\documentclass{article}

\usepackage{spconf,amsmath,epsfig,amssymb,mathrsfs,psfrag,epsf,enumerate,bm}
\usepackage{amstext,amsfonts,amssymb}
\usepackage{float}
\usepackage{graphicx}
\usepackage{cite}
\usepackage{subfigure}
\usepackage{url}
\usepackage{shadow,color,pifont,times,rotate}
\def\esp{{\mathrm{E}}\,}

\newsavebox{\fminibox}
\newlength{\fminilength}

  \def\+{^\dagger}

\def\W{,\thickspace}

\def\nequiv{\not\kern-.05em\equiv}
\def\egal{\kern-.5em=\kern-.5em}        
\def\propt{\kern-.2em\propto\kern-.2em} 

\def\argmin{\mathop{\mathrm{arg\,min}}} 
\def\intdouble{\int\kern-0.3em\int}
\def\inttriple{\int\kern-0.3em\int\kern-0.3em\int}

\def\rond#1{\overset{\kern-0.33em~_\circ}{#1}}
\def\rondit[#1]#2{\overset{\kern#1~_\circ}{#2}}

\def\H{{\mathbf H}}
\def\HH{\widehat{{\mathbf H}}}
\def\tg{\mathcal{H}}
\def\y{{\bm y }}
\def\s{{\bm s }}
\def\z{{\bm z }}
\def\W{{ W}}
\def\WT{{\widetilde{ W}}}
\def\esp{{\mathbb E}}
\def\eye{{\mathbb I}}

\def\sig{\mathbf \Sigma}

\DeclareMathAlphabet{\mathpzc}{OT1}{pzc}{m}{it}

\newcommand{\mb}{\mathbf}

\newcommand{\mc}{\mathcal}

\title{MIMO-OFDM Optimal Decoding And Achievable Information Rates \\Under Imperfect Channel Estimation}

\name{Sajad Sadough$^{\dagger*}$, Pablo Piantanida$^*$, and Pierre Duhamel$^*$}

\address{$^\dagger$Ecole Nationale Sup\'erieure de Techniques Avanc\'ees, 75739 Paris Cedex 15, France\\
$^*$ Laboratoire des Signaux et Syst\`emes, CNRS/Sup\'{e}lec, F-91192 Gif-sur-Yvette, France\\ Email:\{sadough, piantanida, pierre.duhamel\}@lss.supelec.fr \\[1mm]
}
\begin{document}
\ninept
\maketitle
\vspace{-3mm}
\begin{abstract}

Optimal decoding of bit interleaved coded modulation (BICM) MIMO-OFDM where an imperfect channel estimate is available at the receiver is investigated. First, by using a Bayesian approach involving the channel {\it a posteriori} density, we derive a practical decoding metric for general memoryless channels that is robust to the presence of channel estimation errors. Then, we evaluate the outage rates achieved by a decoder that uses our proposed metric.
The performance of the proposed decoder is compared to the classical mismatched decoder and a theoretical decoder defined as the best decoder in the presence of imperfect channel estimation.
Numerical results over Rayleigh block fading MIMO-OFDM channels show that the proposed decoder outperforms mismatched decoding in terms of bit error rate and outage capacity without introducing any additional complexity.
\end{abstract}
\vspace{-5mm}
\section{Introduction}

Multiple-input multiple-output (MIMO) antennas systems is a promising technique for high-speed, spectrally efficient and reliable wireless communications. However, as higher data rates lead to wideband communications, the underlying MIMO channels exhibit strong frequency selectivity. By using orthogonal frequency division multiplexing (OFDM) and applying a proper cyclic prefix (CP), the frequency selective channels are transformed to an equivalent set of frequency-flat subchannels. These considerations motivate the combination of MIMO and OFDM, referred to as MIMO-OFDM, as a promising technology for the future generation of wireless systems \cite{bolskei02}.

It is well known that reliable coherent data detection is not possible unless an accurate channel state information (CSI) is available at the receiver. A typical scenario for wireless communication systems assumes the channel changing so slowly that it can be considered time invariant during the transmission of an entire frame. In such situations, pilot symbol assisted modulation (PSAM) has been shown to be an effective solution for obtaining CSI. It consists of inserting known training symbols (pilot) at the beginning of the information frame from which the receiver estimates the channel before decoding the rest of the frame based on that channel estimate. However, in practical systems, due to the finite number of pilot symbols and noise, the receiver has only access to an imperfect (possibly very bad) estimate of the channel.

In order to deal with imperfect CSI, one sub-optimal technique, known as {\it mismatched} decoding, consists in replacing the exact channel by its estimate in the receiver metric. Although, this scheme is not adapted to the presence of channel estimation errors (CEE), it has been extensively adopted for performance evaluation of single an multi-carrier MIMO systems \cite{garg05}.
However, adopting a mismatched decoding approach arises two important questions: $\textbf{i}$) what are the limits of achievable information rates and how close it is to the rates obtained by a theoretical decoder in the presence of CEE ? and $\textbf{ii}$) what type of practical encoder/decoder can achieve the best performance under imperfect channel estimation ?
The first problem has been addressed in \cite{piantanida} where mismatched decoding has been shown to be largely suboptimal compared to the capacity provided by a theoretical decoder. Since the theoretical encoder/decoder can not be implemented on practical communication systems, we have recently addressed in \cite{sadough06} the second question in the case of multicarrier systems by deriving a practical decoding metric adapted to imperfect channel estimation and its associated achievable rates.\vspace{-0.5mm}

In this work, we address the above problems for more general wideband block fading MIMO-OFDM channels estimated by a finite number of training sequence. In our approach we exploit an interesting feature of PSAM, the availability of CEE statistics, in order to derive the {\it a posteriori} pdf of the perfect channel conditioned on its estimate. The latter let us to formulate a new channel likelihood as a function of the estimated channel and derive our {\it improved} decoding rule adapted to the presence of CEE. Interestingly, the present metric coincides with that proposed in \cite{Biglieri_jour} for MIMO space-time decoding. Furthermore, we evaluate the capacity associated to the improved and mismatched decoders by using the results obtained in \cite{merhav94}. Although in this work we will focus on wideband MIMO-OFDM systems working in a bit interleaved coded modulation (BICM) framework, our results can be applied to general memoryless channels.\vspace{-0.5mm}

The outline of this paper is as follows. In Section \ref{sec:model} we describe the MIMO-OFDM channel and its pilot based estimation. We also calculate the posterior distribution of the perfect channel conditioned on its estimate. This posterior distribution is used in section \ref{sec:metric} to derive the improved decoding metric in the presence of imperfect channel state information at the receiver (CSIR). Section \ref{sec:bicmRX}, provides the application of the improved metric for iterative decoding of BICM MIMO-OFDM. In section \ref{sec:cap}, we evaluate the achievable outage rates of a receiver using the proposed metric.
Section \ref{sec:simul} illustrates via simulations, a comparative performance study of the proposed decoder and section \ref{sec:concl} concludes the paper.\vspace{-0.5mm}

Notational conventions are as follows. Upper and lower case bold symbols are used to denote matrices and vectors, respectively; $\eye_N$ represents an $N \times N$ identity matrix; $\mathbb{E}_{\mathbf{x}}[.]$ refers to expectation
with respect to $\mathbf{x}$; $|.|$ and $\|.\|$ and ${\rm Tr}(.)$ denote matrix determinant, Frobenius norm and matrix trace respectively; $(.)^T$ and $(.)^{\mathcal{H}}$ denote vector transpose and Hermitian transpose, respectively.
\vspace{-3mm}
\section{System Model and Channel Estimation}
\label{sec:model}
\begin{figure}[!t]
\centering
\includegraphics[width=0.46 \textwidth,height=0.15\textheight]{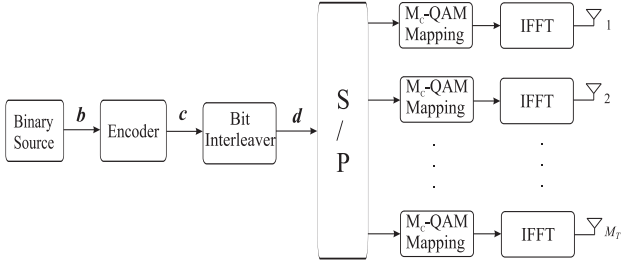}
\caption{Block diagram of MIMO-OFDM transmission scheme.}\label{mimofdm_tx}
\end{figure}

We consider a single-user MIMO-OFDM communication system over a memoryless frequency selective Rayleigh fading channel. The system consists of $M_T$ transmit and $M_R$ receive antennas ($M_R \geq M_T$), and $M$ is the total number of subcarriers. Fig. \ref{mimofdm_tx} depicts the BICM coding scheme used at the transmitter. The binary data sequence ${\bm b}$ are encoded by a non-recursive non-systematic convolutional (NRNSC) code before being interleaved by a quasi-random interleaver. The output bits ${\bm d}$ are multiplexed to $M_T$ substreams and mapped to complex ${\rm M_c}$-QAM symbols before being modulated by the OFDM modulator and transmitted through $M_T$ antennas.

Let $\s$ be the $M M_T\times 1$ vector containing the OFDM symbols transmitted simultaneously over $M_T$ antennas. The symbols are assumed to be independent identically distributed (i.i.d.) with zero mean and unit covariance matrix $\sig_s=\esp[\s\s^\tg]=\eye_{M\times{M_T}}$. Assuming an invariant channel over a frame of $L$ symbols, the received vector $\y$ at a given time index $l$ (omitted for brevity) can be written as
\begin{equation}
\label{eq:model1}
\y = \H \, \s + \z
\end{equation}
where $\H$ is a $M M_R \times M M_T$ block diagonal channel matrix containing the frequency response of the MIMO channels and the noise vector $\z$ is assumed to be a zero-mean circularly symmetric complex Gaussian  (ZMCSCG) random vector with covariance matrix $\sig_z \triangleq \mathbb{E}(\z \z^{\mathcal{H}})=\sigma^2_z \mathbb{I}_{M\times{M_R}}$. We assume that for each frame, a different realization of $\H$ independent of both $\s$ and $\z$ is drawn and remains constant during this frame.
The MIMO-OFDM channel can be decoupled into $M$ frequency flat MIMO channels by exploiting the block diagonal structure in \eqref{eq:model1} which can be rewritten as a set of $M$ equations that contains only one subcarrier each
\begin{equation}
\label{eq:model2}
\y_k = \H_k \, \s_k + \z_k \;\;\;\; k=1,...,M,
\end{equation}
where $\H = {\rm diag}{[\H_1 \H_2 \,...\, \H_M]}$, $\y^T=[\y_1^T \, ...\, \y_M^T]$, $\s^T=[\s_1^T \, ...\, \s_M^T]$ and $\z^T=[\z_1^T \, ...\, \z_M^T]$. The architecture of \eqref{eq:model2} constitutes the basis for the study in this paper.
\vspace{3mm}
\\
\textbf{Pilot Based Channel Estimation:}
We consider the estimation of channel matrix $\H_k$ via the transmission of $N$ training vectors $\s_{_{T,i}}$,   ($i=1,...,N$).
According to \eqref{eq:model2}, when pilot symbols are transmitted we receive
 \begin{equation}
   \label{eq:model3}
         \mathbf{Y}_T = \H_k \,\mathbf{S}_T + \mathbf{Z}_T
  \end{equation}
where each column of the $M_T \times N$ matrix $\mathbf{S}_T=[\s_{_{T,1}}|...|\s_{_{T,N}}]$ contains one pilot symbol and the noise $\mathbf{Z}_T$ has the same distribution as the noise $\z_k$. The average energy of the training symbols is $P_T = \frac{1}{N M_T}{\rm Tr}\big(\mb{S}_T \mb{S}_T^\tg\big)$.
The ML estimate of $\H_k$ is obtained by minimizing $\|\mathbf{Y}_T-\H_k \,\mathbf{S}_T\|^2$ with respect to $\H_k$. We have
\begin{equation}
\label{eq:model4}
\HH_k^{\rm ML}=\mathbf{Y}_T \,\mathbf{S}_T^\tg\,(\mathbf{S}_T\mathbf{S}_T^\tg)^{-1}=\H_k + \mathbf{\mathcal{E}}
\end{equation}
where $\mathbf{\mathcal{E}}= \mathbf{Z}_T \mathbf{S}_T^\tg\,(\mathbf{S}_T\mathbf{S}_T^\tg)^{-1}$ denotes the estimation error matrix. When the training sequence is orthogonal ($\mathbf{S}_T \mathbf{S}_T^\tg = N P_T \eye_{M_T}$), the $j$-th column $\mathbf{\mathcal{E}}
_j$ of the estimation error matrix reduces to a white noise vector with covariance matrix $\sig_{\mathcal{E}}= \esp\big[\boldsymbol{\mathcal{E}}_j \boldsymbol{\mathcal{E}}_j^\tg\big]=\sigma_{\mathcal{E},k}^2 \eye_{M_R}$ (for all $j$), where $\sigma_{\mathcal{E},k}^2=\frac{\sigma^2_z}{N P_T}$.
By assuming that the channel matrix $\H_k$ has the {\it prior} distribution $\H_k \sim \mathcal{CN}(\mathbf{0},\eye_{M_T} \otimes \sig_{H,k})$, ($\sig_{H,k}=\sigma^2_h\, \eye_{M_R}$) and choosing an orthogonal training sequence, we can derive the posterior distribution of the perfect channel conditioned on the estimated channel as
\begin{equation}
\label{eq:model5}
f(\H_k|\HH_k^{\rm ML}) = \mathcal{CN}( \sig_{\Delta} \HH_{\rm ML} ,\, \eye_{M_T} \otimes \sig_{\Delta} \sig_{ \mathcal{E} } )
\end{equation}
where $\sig_{\Delta}=\sig_{H,k}(\sig_{\mathcal{E}}+\sig_{H,k})^{-1}=\delta \eye_{M_R}$ and $\delta=\frac{ \sigma^2_h}{\sigma^2_h+\sigma^2_{\mathcal{E},k}}$.

The availability of the estimation error distribution constitutes an interesting feature of pilot assisted channel estimation that we used to derive the posterior distribution \eqref{eq:model5}. This distribution is exploited in the next section, in the derivation of a new metric for improving the detection performance under imperfect channel estimation. \vspace{-2mm}
\section{Maximum-Likelihood Detection in the Presence of Channel Estimation Errors}
\label{sec:metric}
\vspace{-2mm}
\subsection{Mismatched ML Detection}
\label{subsec:mis}
It is well known that under i.i.d. Gaussian noise, detecting $\s_k$ is given by maximizing the likelihood function $\W(\y_k|\s_k,\H_k)$ which is equivalent to minimizing the Euclidean distance $\mathcal{D}_{\rm ML}$
\begin{equation}
\label{eq:metric1}
\hat{\s}_k^{\rm ML}(\H_k)= \argmin_{\s_k \in \, \mathbb{C}^{M_T \times 1}}\, \big \{ \,\mathcal{D}_{\rm ML}(\s_k,\y_k,\H_k) \big \},
\end{equation}
where
$\mathcal{D}_{\rm ML}(\s_k,\y_k,\H_k) \triangleq - \ln \W(\y_k|\s_k,\H_k)\propto \|\y_k - \H_k \s_k \|^2$.
Since the above detection rule requires the knowledge of the {\it perfect} channel matrix $\H_k$, one sub-optimal approach, referred to as mismatched detection, consists in replacing the exact channel by its estimate in the receiver metric as
\begin{equation}
\label{eq:metric3}
\hat{\s}_k^{\rm ML}(\HH_k)= \argmin_{\s_k \in \, \mathbb{C}^{M_T \times 1}}\, \big \{ \,\|\y_k - \H_k \s_k \|^2 \big \}_{\big |_{\H_k=\HH_k} }
\end{equation}
\vspace{-4mm}
\subsection{Improved ML Detection for imperfect CSIR}
\label{subsec:mod}
The Bayesian framework introduced previously let us to define a new likelihood function $\WT(\y_k|\HH_k,\s_k)$ by averaging the likelihood that would be used if the channel were perfectly known ($W(\y_k|H_k,\s_k)$), over all realizations of the perfect channel for a given estimated channel state (the posterior distribution \eqref{eq:model5}). This yields
\begin{align}
	\label{eq:metric4}
             \WT(\y_k|\HH_k,\s_k) &= \esp_{\H_k | \HH_k} \big[ \W(\y_k|\H_k,\s_k) \big | \, \HH_k \big] \notag \\
             &= \int_{\H_k \in \mathbb{C}^{M_R \times M_T}} \W(\y_k|\H_k,\s_k) \; f(\H_k|\HH_k) \;\;\; {\rm d}\H_k.
\end{align}
Since both $\W(\y_k|\H_k,\s_k)$ and $f(\H_k|\HH_k)$ are gaussian densities, it is easy to show that
$\WT(\y_k|\HH_k,\s_k) = \mathcal{CN}(\boldsymbol{\mu}_{_\mathcal{M}},\sig_{_\mathcal{M}})$                              with \cite{sadough06}
\begin{equation}
  \label{eq:metric5}
  \left\{\begin{array}{ll}
       \boldsymbol{\mu}_{_\mathcal{M}} &= \delta \, \HH_k \, \s_k, \\
       \sig_{_\mathcal{M}} &= \sig_z  + \delta \, \sig_{\mathcal{E}} \, \| \s_k \|^2.
    \end{array} \right.
\end{equation}
Now the ML estimate of $\s_k$ can be formulated as
\begin{equation}
\label{eq:metric6}
  \hat{\s}_k^{\mathcal{M}}(\HH_k)= \argmin_{\s_k \in \, \mathbb{C}^{M_T \times 1}}\, \big \{ \,\mathcal{D}_{\mathcal{M}}(\s_k,\y_k,\H_k) \big \},
\end{equation}
with
\begin{align}
\label{eq:metric7}
& \mathcal{D}_{_\mathcal{M}}(\s_k,\y_k,\HH_k) \triangleq - \ln \WT(\y_k|\s_k,\HH_k) \notag \\
&= M_R \ln \, \pi (\sigma^2_z + \delta \, \sigma^2_\mathcal{E}\, \|\s_k\|^2)+ \frac{\|\y_k-\delta\,\HH_k\,\s_k\|^2}{\sigma^2_z+\delta\,\sigma^2_\mathcal{E}\,\|\s_k\|^2}
\end{align}
being the new ML decision metric under CEE.

We note that when exact channel is available ($\HH_k=\H_k$), the posterior expectation of \eqref{eq:metric4} becomes equivalent to replacing $\H_k$ by $\HH_k$ in $\W(\y_k|\H_k,\s_k)$ and consequently the the two metrics $\mathcal{D}_{_\mathcal{M}}$ and $\mathcal{D}_{\rm ML}$ coincides. Under near perfect CSIR, occurred either when $\sigma^2_{\mathcal{E}} \to 0$ or when $N \to \infty$, we have $\delta \to 1$, $\delta \sigma^2_{\mathcal{E}} \to 0$,
$\boldsymbol{\mu}_{_\mathcal{M}} \to \HH_k \, \s$ and $\sig_{_\mathcal{M}} \to \sigma^2_z \eye_{M_R}$.
Consequently, the improved metric $\mathcal{D}_{_\mathcal{M}}$ behaves similarly to the classical Euclidean distance metric $\mathcal{D}_{\rm ML}$.
\begin{figure}[!t]
\centering
\includegraphics[width=0.46\textwidth,height=0.3\textheight]{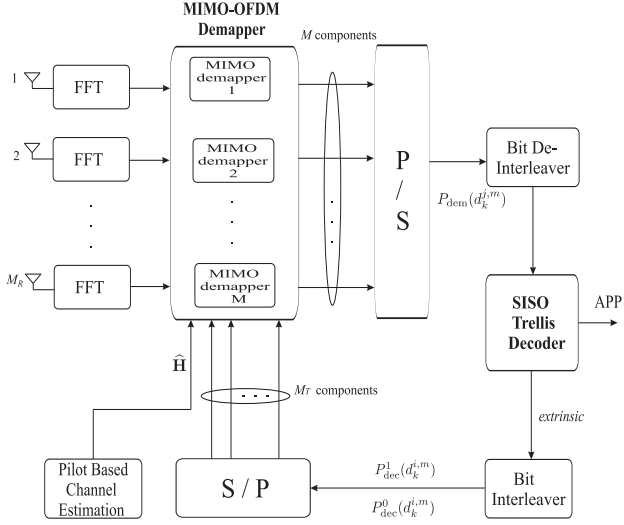}
\caption{Block diagram of MIMO-OFDM BICM receiver.}\label{mimofdm_rx}
\end{figure}
\vspace{-3mm}
\section{Iterative Decoding of BICM MIMO-OFDM Based On Imperfect CSIR}
\label{sec:bicmRX}

As a practical application of the general decoding metric \eqref{eq:metric7}, we consider soft iterative decoding of BICM MIMO-OFDM under imperfect CSIR. This problem has been addressed in \cite{boutros00} under the assumption of perfect CSIR. Here, without going into the details, we extend the results of \cite{sadough06} to MIMO-OFDM block fading channels estimated by a finite number of training symbols.

As shown in Fig. \ref{mimofdm_rx}, the BICM receiver consists of a bunch of demodulator/demapper, a de-interleaver
and a soft-input-soft-output (SISO) decoder. Let $d_k^{i,m}$ be the $m$-th coded and interleaved bit ($m=1,2,...,B=\log_2 {\rm M_c}$) of the constellation symbol $\s_k$ at the the $i$-th transmit antenna and the $k$-th subcarrier.
We denote by $L(d_k^{i,m})$ the coded log-likelihood ratio (LLR) value of the bit $d_k^{i,m}$.
At each decoding iteration, the LLR values conditioned on the CSIR are given by
\begin{equation}
\label{eq:mimoRx1}
L(d_k^{i,m})=\log \frac{P_{\rm dem}(d_k^{i,m}= 1)|\y_k,\H_k)}{P_{\rm dem}(d_k^{i,m}=0|\y_k,\H_k)}.
\end{equation}
We have (see \cite{sadough06} and references therein)
\begin{equation}
\label{eq:mimoRx2}
L(d_k^{i,m})= \log \frac{\sum \limits_{\s_k:d_k^{i,m}=1} \W(\y_k|\s_k,\H_k) \prod \limits_{\stackrel{n=1}{n\neq m}}^{B M_T} P^1_{\rm dec}(d_k^{i,n})}{\sum \limits_{\s_k:d_k^{i,m}=0} \W(\y_k|\s_k,\H_k) \prod \limits_{\stackrel{n=1}{n\neq m}}^{B M_T} P^0_{\rm dec}(d_k^{i,n})}\vspace{-2mm}
\end{equation}
where $P^1_{\rm dec}(d_k^{i,m})$ and $P^0_{\rm dec}(d_k^{i,m})$ are {\it extrinsic} information coming from the SISO decoder.

Notice that the channel likelihood $\W(\y_k|\s_k,\H_k)$ involved in \eqref{eq:mimoRx2} is conditioned on the {\it perfect} channel $\H_k$ of which the receiver has {\it solely} an estimate $\HH_k$.
In order to enhance the detection performance under imperfect CSIR, we propose an {\it improved} decoder that uses $\WT(\y_k|\s_k,\HH_k)$ of \eqref{eq:metric4} instead of $\W(\y_k|\s_k,\HH_k)$ (mismatched approach), for the derivation of the coded LLRs of \eqref{eq:mimoRx2}. The decoder accepts the LLRs
of all coded bits and employs the well known forward-backward algorithm \cite{bcjr} to compute the LLRs of information bits, which are used for the decision.
\vspace{-3mm}
\section{Achievable Rates of MIMO-OFDM Under Channel Estimation Errors}
\label{sec:cap}
In this section we provide the instantaneous achievable information rates $C_{\mathcal{M}}$ and $C_{\rm ML}$ associated to a receiver using the decoding rules of \eqref{eq:metric6} (improved) and \eqref{eq:metric3} (mismatched), respectively. We consider a MIMO-OFDM channel $W(\y|\s,\H)=\prod_{k=1}^M \mc{CN}(\H_k \s_k,\sig_{\z_k})$, with $\sig_{\z_k}=\sigma^2_{z,k}\,\eye_{M_R}$. Furthermore, we assume that the transmitter does not disposes of the channel estimate to perform power control. Thus, equal power is allocated to symbols $\s_k$ by setting $\sig_{\s_k}=\bar{P}\, \eye_{M_T}$.
\subsection{Achievable Rates Associated to the Improved Decoder }
The capacity of a general mismatched decoder assuming Gaussian i.i.d inputs $\s \sim \mc{CN}(\mathbf{0},\sig_s)$ is given in \cite{merhav94} by minimizing the mutual information $I(S;Y)$ defined for MIMO-OFDM channels as \cite{bolskei02}
\begin{equation}
\label{mutinf}
I(S;Y)\triangleq \frac{1}{M} \sum_{k=1}^M \log_2 \det \left(\eye_{M_R}+ \Upsilon_k \sig_{\s_k} \Upsilon_k^{\tg}\sig_k^{-1} \right)
\end{equation}
where the minimization is done over all general channels $V(\y|\s,\Upsilon)=\prod_{k=1}^M \mc{CN}(\Upsilon_k \s_k,\sig_k)$.
Since in the present work the decoding is performed on a single subcarrier basis, the minimization can be done independently for each subcarrier according to its corresponding channel and estimate pair $(\H_k,\HH_k)$.
Thus, we have to find the optimal $\Upsilon_k\in \mathbb{C}^{M_R\times M_T}$ and covariance matrices $\Sigma_k = \sigma^2_k\mathbb{I}_{M_R}$ so as to minimize $I(S;Y)$ under the set of constraints \cite{merhav94} \vspace{2mm}

$({\rm c_{1}}):\rm{Tr}\big(\mathbb{E}_{\s}[V(\y|\s,\Upsilon)]\big)=\rm{Tr} \big(\mathbb{E}_{\s}[W(\y|\s,\H)]\big)$, \vspace{2mm}

$({\rm c_{2,k}}): \mathbb{E}_{\s_k} \Big [\mathbb{E}_{V}[\mathcal{D}_\mc{M}(\s_k,\y_k|\HH_k)] \Big] \leq \\
\mathbb{E}_{\s_k} \Big[\mathbb{E}_{W} \big[\mathcal{D}_\mc{M}(\s_k,\y_k|\HH_k)\big]\Big]$,
\\
for $k=1,...,M$.
Applying our channel model \eqref{eq:model1}, and after some calculus, the above set of two constraints are obtained as 
\begin{align}
&\mathrm{(c_{1})}: \det \left(\Upsilon \sig_\s \Upsilon^\tg + \sig \right)=\det\left(\H \sig_\s \H^\tg + \sig_{\z}\right) \label{c1}\\
&\mathrm{(c_{2,k})}: \| \Upsilon_k + a_k \widehat{\H}_k \|^2\leq  \| \H_k +a_k \widehat{\H}_k \|^2 + \textrm{Cst}_k,\label{c2}
\end{align}
where
\\
$a_k=\delta_k(\delta_k \sigma^2_{\mathcal{E},k} \bar{P} - \lambda_{n,k} \sigma_{z,k}^2)\big[M_T\delta\sigma^2_{\mc{E},k}\lambda_{n,k} \bar{P} +\lambda_{n,k} \sigma_{z,k}^2- \delta_k \sigma^2_{\mathcal{E},k} \bar{P} \big]^{-1}$;\vspace{2mm}
$\textrm{Cst}_k = M_T \lambda_{n,k} \big[\|\H_k \|^2-\| \Upsilon_k\|^2 + \frac{1}{\bar{P}}\big({\rm Tr}(\sig_{\z_k})-{\rm Tr}(\sig_k)\big)\big] \big[1-\frac{\sigma_{z,k}^2}{\delta_k \bar{P}\sigma^2_{\mathcal{E},k}}\lambda_{n,k}-M_T \lambda_{n,k} \big]^{-1}$,
and \vspace{2mm}
\\
$\lambda_{n,k}=\left(\frac{ \sigma_{z,k}^2}{\delta_k \bar{P}\sigma^2_{\mathcal{E},k} }\right)^n\exp\left(\frac{ \sigma_{Z,k}^2}{\delta_k \bar{P} \sigma^2_{\mathcal{E},k} }\right)\Gamma\left(-n,\frac{ \sigma_{Z,k}^2}{\delta_k \bar{P} \sigma^2_{\mathcal{E},k}}\right)$,
\\
with $n=M_T-1$, and
\begin{equation*}
\Gamma (-n,t)=\frac{(-1)^n}{n !} \Big[\Gamma (0,t)-\exp(-t)\sum\limits_{i=0}^{n-1}(-1)^i \frac{i !}{t^{i+1}} \Big], \vspace{-1mm}
\end{equation*}
where $\Gamma (0,t)$ is the exponential integral function.

We consider the singular value decomposition (SVD) of $\H_k=\mathbf{U}_k \Lambda_k \mathbf{V}_k^\tg$ with $\Lambda_k=\textrm{diag}(\lambda_{k,1},\dots,\lambda_{k,M_R})$. Let $D_{\mu,k}$ be a diagonal matrix such that $D_{\mu,k}=\mathbf{U}_k^\tg \Upsilon_k \mathbf{V}_k$, whose diagonal values are given by the vector $\underline{\mu}_k=(\mu_{k,1},\dots,\mu_{k,M_R})^T$. We define $\widetilde{\H}_k^\tg=\mathbf{V}_k^\tg \widehat{\H}_k^\tg \mathbf{U}_k$ and the vector $\mb{\tilde{h}}_k^\tg=\textrm{diag}(\widetilde{\H}_k^\tg)^T$ resulting of its diagonal and let $b_k= \| \H_k + a_k \widehat{\H}_k\|^2 - a_k^2 (\| \widetilde{\H}_k\|^2-\|  \mb{\tilde{h}}_k\|^2)$. Using the above definitions, it is easy to reformulate \eqref{mutinf} as
\begin{equation}
C_{\mc{M}}(\H,\widehat{\H})=\left \{ \begin{array}{ll} \min \limits_{\underline{\mu}} \,\,\,\,\,\,  \frac{1}{M}\sum\limits_{k=1}^{M} \sum\limits_{i=1}^{M_R}\log_2 \left(1+\displaystyle{\frac{\bar{P}|\mu_{k,i}|^2}{\sigma^2( \underline{\mu_k})}}\right), \\ \textrm{subject to} \,\,\,\,\,\,  \| \underline{\mu}_k +a  \mb{\tilde{h}_k}  \|^2\leq b_k. \end{array}\right.\label{final_opt}
\end{equation}
According to \eqref{final_opt}, for each subcarrier $k=1,...,M$, the achievable rates associated to the metric $\mathcal{D}_{\mc M}$ of \eqref{eq:metric6}, can be obtained by using standard Lagrange method. We have
\begin{equation}
C_{\mc{M}}(\H,\widehat{\H})=\frac{1}{M}\sum\limits_{k=1}^M\log_2 \textrm{det}\left(\mathbb{I}_{M_R}+ \frac{\bar{P} \Upsilon_{\textrm{opt},k}\Upsilon_{\textrm{opt},k}^\tg}{ \sigma_k^{2}( \underline{\mu}^{\textrm{opt}}_{\mc{M},k})}   \right),\label{acievable_rates}
\end{equation}
where $\sigma_k^{2}(\underline{\mu}^{\textrm{opt}}_{\mc{M},k})=\frac{\bar{P}}{M_T}(\|\Lambda_k \|^2-\|D_{\mu,k} \|^2)+\sigma_{z,k}^2$,
\\
$\Upsilon_{\textrm{opt},k}=\mathbf{U}_k D_{\mu^{\textrm{opt}}_{\mc{M},k}} \mathbf{V}_k^\tg$ and the vector containing the diagonal elements of the optimal solution $D_{\mu^{\textrm{opt}}_{\mc{M},k}}$ is given by
\begin{equation}
\underline{\mu}^{\textrm{opt}}_{\mc{M},k}=\left\{ \begin{array}{ll} \displaystyle{\left(\frac{\sqrt{b_k}}{\|\mb{\tilde{h}}_k\|}-|a_k|\right)\mb{\widetilde{h}}_k} & \,\, \textrm{if $b_k\geq 0$,} \\
  \underline{0} & \,\, \textrm{otherwise}. \end{array}  \right. \label{solution_mu}
\end{equation}

Similarly, we can compute the achievable rates associated to the mismatched ML decoder \eqref{eq:metric3}. This is given by replacing the vector $\underline{\mu}^{\textrm{opt}}_{\mc{M},k}$ by \vspace{-2mm}
\begin{equation}
\label{mismached_ML_capa}
\underline{\mu}^{\textrm{opt}}_{\textrm{ML},k}=\frac{ \mathbb{R}e\{{\rm Tr}(\Lambda_k^\tg \tilde{\mb{h}}_k) \}}{\|\tilde{\mb{h}}_k\|^2}\tilde{\mb{h}}_k,
\end{equation}
in equation \eqref{acievable_rates}.
\subsection{Outage Rates Evaluation}
Given any pair of matrices $(\H,\HH)$, the expression \eqref{acievable_rates} provides the instantaneous achievable rates associated to a receiver using the decoding rule \eqref{eq:metric6}. The outage probability associated to an outage rate $R \geq 0$ is defined as
\begin{equation}
P_{_{\mathcal{M}}}^{\mathrm{out}}(R,\widehat{\mathbf{H}})=\mathrm{Pr}_{\mathbf{H}| \widehat{\mathbf{H}}} \big(\mathbf{H}\in \Lambda_{_\mathcal{M}}(R,\widehat{\mathbf{H}})|\widehat{\mathbf{H}}\big), \notag
\end{equation}
with $\Lambda_{_\mathcal{M}}(R,\widehat{\mathbf{H}})=\big\{\mathbf{H}:\, C_{_\mathcal{M}}(\mathbf{H},\widehat{\mathbf{H}})<R\big\}$. Using this, the outage rate of the improved decoder for an outage probability $\gamma$ is
\begin{equation}
\label{capout}
C_{\mathcal{M}}^{\mathrm{out}}(\gamma,\widehat{\mathbf{H}})=\sup_R\big\{R \geq 0: P_{_{\mathcal{M}}}^{\mathrm{out}}(R,\widehat{\mathbf{H}})\leq \gamma\big\}
\end{equation}
and $C_{\textrm{ML}}^{\mathrm{out}}(\gamma,\widehat{\mathbf{H}})$ (for mismatched ML decoder), is given by replacing $\Lambda_{_\mathcal{M}}(R,\widehat{\mathbf{H}})$ by $\Lambda_{\textrm{ML}}(R,\widehat{\mathbf{H}})=\big\{\mathbf{H}:\, C_{\textrm{ML}}(\mathbf{H},\widehat{\mathbf{H}})<R\big\}$ in equation \eqref{capout}.
Since these outage rates still depend on the random channel estimate $\widehat{\mathbf{H}}$, we will consider the expected outage rates over all channel estimates as
\begin{equation}
\overline{C}_{\mathcal{M}}^{\; \mathrm{out}}(\gamma)=\mathbb{E}_{\widehat{\mathbf{H}}}\big[C_{_\mathcal{M}}^{\mathrm{out}}(\gamma,\widehat{\mathbf{H}})\big]. \label{EQ-capacity}
\end{equation}
The achievable rates \eqref{EQ-capacity} are upper bounded by the outage rates provided by a theoretical decoder (i.e. the best decoder in the presence of CEE). In our case, this capacity is given by
\begin{equation}
\overline{C}_{_\mathcal{G}}^{\; \mathrm{out}}(\gamma)=\mathbb{E}_{\widehat{\mathbf{H}}}\big[C_{_\mathcal{G}}^{\mathrm{out}}(\gamma,\widehat{\mathbf{H}})\big], \label{perfect-capacity}
\end{equation}
where the outage rates $C_{_\mathcal{G}}^{\mathrm{out}}(\gamma,\widehat{\mathbf{H}})$ are computed using
\begin{equation}
C_{_\mathcal{G}}(\mathbf{H},\widehat{\mathbf{H}})=\frac{1}{M}\sum\limits_{k=1}^{M} \log_2 \det \left(\eye_{M_R} +\frac{\bar{P}\;\H_k\H_k^\tg}{\sigma^2_z}\right),
\end{equation}
and $\H_k$ are random channels drawn from the posterior distribution of \eqref{eq:model5}.
\vspace{-3mm}
\section{Numerical Results}
\label{sec:simul}

Next, the performance of the improved decoder is measured in terms of bit error rate (BER) and achievable outage rates. The binary information data are encoded by a rate $1/2$ NRNSC code with constraint length 3 defined in octal form by (5,7). Throughout the simulations, each frame is assumed to consists of one OFDM symbol with 50 subcarriers belonging to a 16-QAM constellation with Gray labeling. The interleaver is a pseudo-random one operating over the entire frame with size $M\! \cdot\! M_T\! \cdot\! \log_2(\rm{M_c})$ bits.
For each transmitted frame, a different realization of a Rayleigh distributed channel has been drawn and remains constant during the whole frame. Besides, it is assumed that the average pilot symbol energy is equal to the average data symbol energy. Moreover, the number of decoding iterations are set to 4.

\begin{figure}[!htb]
\vspace{-3mm}
\centering
\includegraphics[width=0.5\textwidth,height=0.29\textheight]{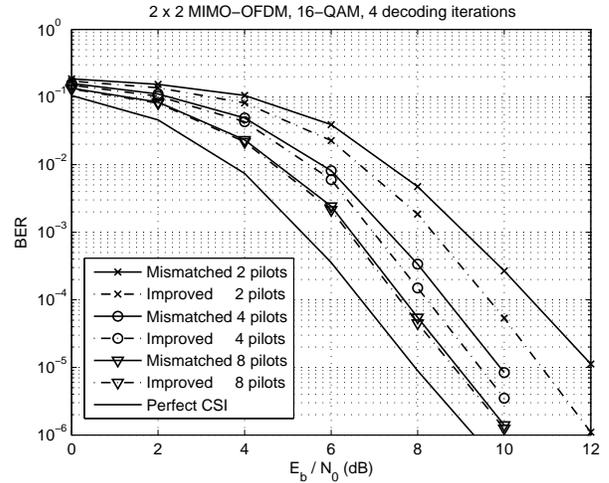}
\caption{BER performance of the proposed and mismatched decoders in the case of a $2\times2$ MIMO-OFDM Rayleigh fading channel with $M=50$ subcarriers for training sequence length $N\in\{2, 4, 8\}$.} \label{bermimofdm} \vspace{-4mm}
\end{figure}

Fig. \ref{bermimofdm} depicts the BER performance over a $2\times2$ MIMO-OFDM channel estimated by $N\in\{2,4,8\}$ pilot symbols per frame. As observed, the increase in the required $E_b/N_0$ caused by CEE is an
important effect of imperfect CSIR in the case of mismatched ML decoding.
The figure shows that the SNR to obtain a BER of $10^{-5}$ with $N=2$ pilots is reduced by about $1.3$ dB if the improved decoder is used instead of the mismatched decoder. We also notice that the performance loss of the mismatched receiver with respect to the derived receiver becomes insignificant for $N\geq8$. This can be explained from the expression of the metric \eqref{eq:metric7}, where we note that by increasing the number of pilot symbols, this expression tends to the classical Euclidean distance metric. However, the proposed decoder outperforms the mismatched decoder especially when few numbers of pilot symbols are dedicated for channel estimation.

\begin{figure}[!htb]
\vspace{2mm}
\centering
\includegraphics[width=0.5\textwidth,height=0.28\textheight]{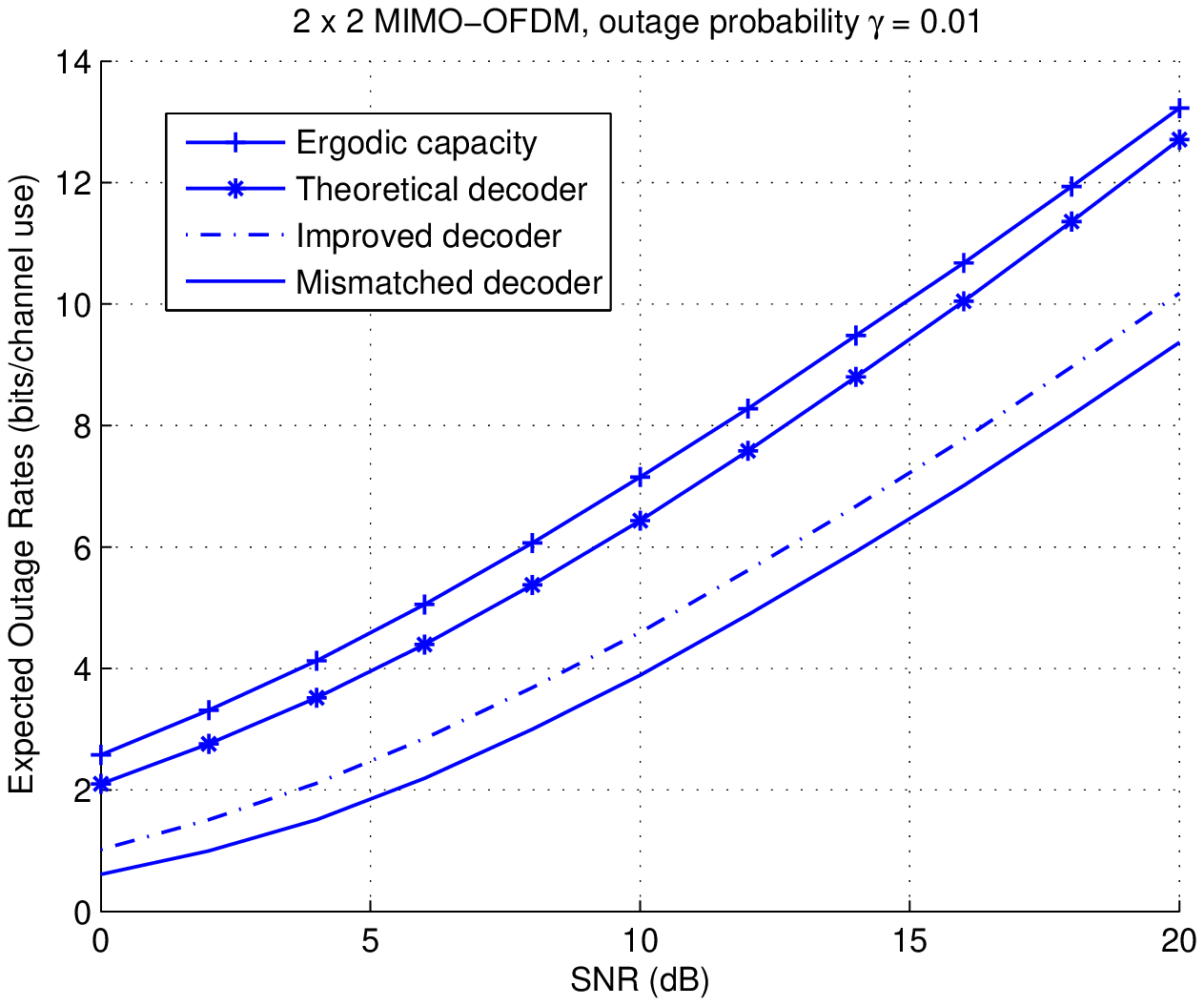}
\caption{Expected outage rates of a $2\times2$ MIMO-OFDM system with $M=16$ subcarriers versus SNR for $N=2$ pilots.}\label{capmimofdm22}\vspace{-4mm}
\end{figure}

Fig. \ref{capmimofdm22} shows the expected outage rates (in bits per channel use) corresponding to the transmission of one OFDM symbol with $M=16$ subcarriers and $M_T=M_R=2$ antennas, achieved by adopting mismatched ML decoding and the improved decoder (expression \eqref{EQ-capacity}). For comparison, we also display the upper bounds on these achievable outage rates (expression \eqref{perfect-capacity}) and the ergodic capacity.
$N=2$ pilot symbols are sent per frame for CSIR acquisition and the outage probability has been fixed to $\gamma=0.01$. At a mean outage rate of $8$ bits, we note that the achievable rate of the mismatched decoder is about $5$ dB of SNR far from the capacity provided by the theoretical decoder. As observed, by using the improved decoder, higher rates are obtained for any SNR values and the aforementioned SNR gap is reduced by about $1.8$ dB.

\begin{figure}[!htb]
\vspace{2mm}
\centering
\includegraphics[width=0.5\textwidth,height=0.28\textheight]{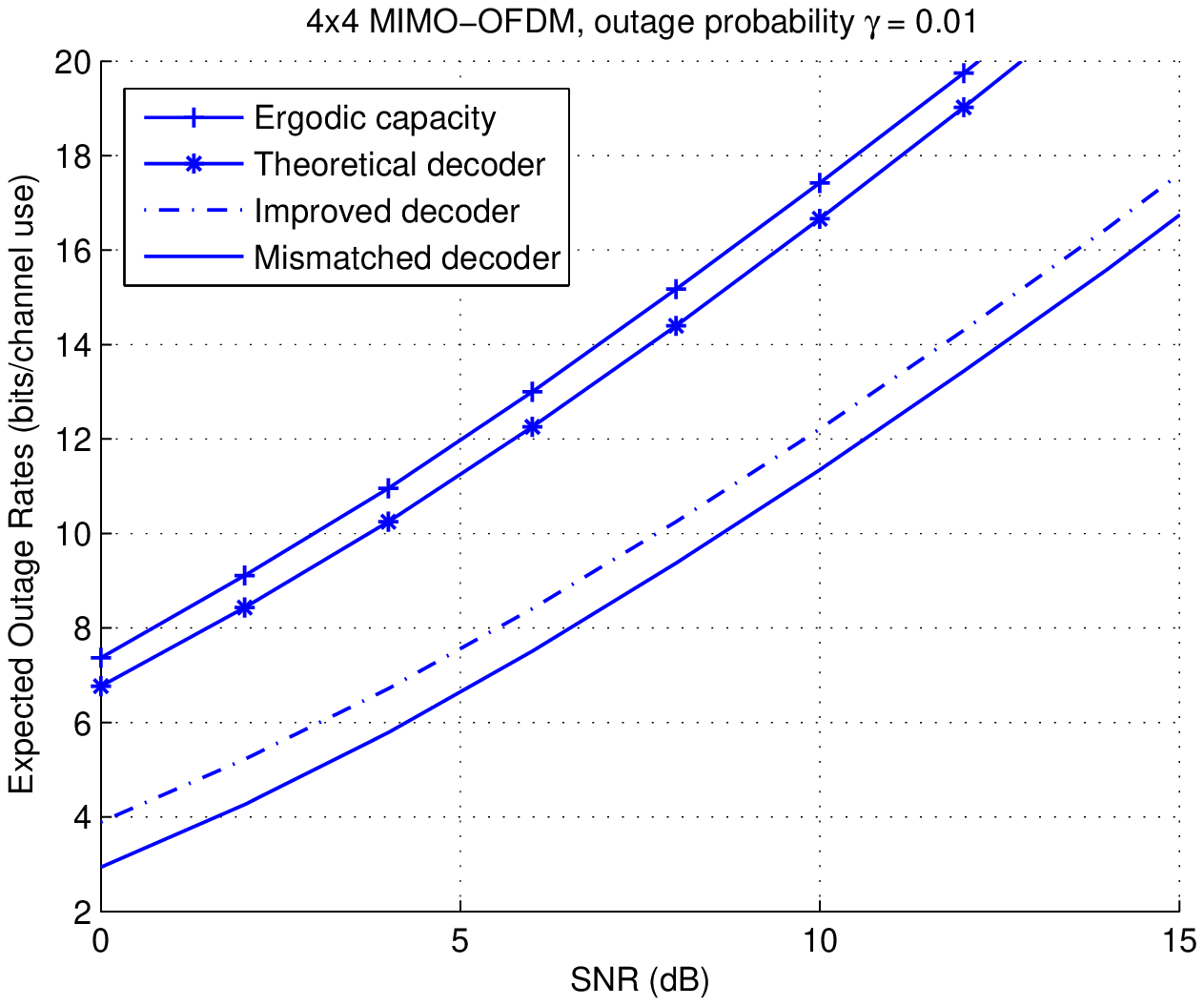}
\caption{Expected outage rates of a $4\times4$ MIMO-OFDM system with $M=16$ subcarriers versus SNR for $N=4$ pilots.}\label{capmimofdm44}\vspace{-4mm}
\end{figure}

Similarly, Fig. \ref{capmimofdm44} compares the outage rates in the case of a $4\times4$ MIMO-OFDM channel estimated by $N=4$ training symbols. Again, it can be observed that the proposed decoder achieves higher rates than the mismatched decoder. However, we note that the increase in the SNR (at a given mean outage rate) induced by using the mismatched decoder rather than the improved decoder, is less than that obtained for a $2\times2$ MIMO-OFDM channel (see Fig. \ref{capmimofdm22}). This can be explained by noting that when $M_T=4$, we must send $N \geq M_T=4$ pilot symbols per frame which yields a more accurate estimate of the channel and consequently bring closer the performance of the two decoders. This observation is in consistence with those presented in \cite{garg05} where it is reported that the performance degradation due to imperfect channel estimation can be reduced by increasing the number of antennas.
\vspace{-1.5mm}
\section{Conclusion}
\label{sec:concl}
A Bayesian approach for the design of a decoding method that is robust to channel estimation errors was presented.
The robustness of our method comes from the averaging of the decoding rule, that would be used if the channel were perfectly known, over all channel estimation errors.
We also derived the expression of the achievable rates associated to our proposed decoder and compared it to that provided by the classical mismatched decoding approach. As a practical application, the proposed decoder was exploited for iterative BICM decoding of MIMO-OFDM under imperfect channel knowledge.
Simulation results showed that, without introducing any additional complexity, the proposed decoder outperforms the classical mismatched approach in terms of BER and achievable outage rates for short training sequences.
Although our proposed method outperforms mismatched decoding, the derivation of a practical decoder achieving the maximum outage rate under imperfect channel estimation is still an open problem.
\vspace{-1.5mm}
\bibliographystyle{IEEEbib}
\bibliography{biblio}
\end{document}